\documentclass[a4paper]{easychair}
\usepackage{latexsym}
\usepackage{amssymb,amsmath}
\usepackage{fancybox}
\usepackage{color}
\usepackage{comment}
\usepackage{graphics}
\usepackage{epsfig}
\usepackage{multicol}
\usepackage{multirow}
\usepackage{wrapfig}
\usepackage{subcaption}
\newtheorem{definition}{Definition}

\newtheorem{example}{Example}

\newcommand{\A}{{\mathcal A}}
\newcommand{\C}{{\mathcal C}}
\newcommand{\R}{{\mathcal R}}
\renewcommand{\P}{{\mathcal P}}
\newcommand{\grant}{{\sf grant}}
\newcommand{\deny}{{\sf deny}}

    {\end{array}$$}
\newcommand{\ignore}[1]{}

\newcommand{\CBAC}{\mbox{CBAC}}

\newcommand{\PAR}{\mathcal{PAR}}
\newcommand{\ARCA}{\mathcal{ARCA}}
\newcommand{\PCA}{\mathcal{PCA}}
\newcommand{\BAR}{\mathcal{BAR}}
\newcommand{\BARCA}{\mathcal{BARCA}}
\newcommand{\UNDET}{\mathcal{UNDET}}
\newcommand{\CBACO}{\mbox{CBACO}}

\newcommand{\RBAC}{\mbox{RBAC}}

\newcommand{\E}{\mathcal{E}}
\newcommand{\GE}{\mathcal{G}}
\renewcommand{\O}{\mathcal{O}}
\newcommand{\D}{\mathcal{D}}
\newcommand{\oblOr}{\subseteq_{\mathcal{O}}}
\renewcommand{\H}{\mathcal{H}}
\newcommand{\OARCA}{\mathcal{OCA}}
\newcommand{\OPAR}{\mathcal{OPA}}
\newcommand{\DPAR}{\mathcal{DA}}
\newcommand{\ET}{\mathcal{ET}}
\newcommand{\EI}{\mathcal{EI}}
\newcommand{\gspec}{\spec^{\emptyset}}
\newcommand{\spec}{\mathsf{spec}}
\newcommand{\gv}{\v^{\emptyset}}
\renewcommand{\v}{v}

\newcommand{\sign}{\Delta}
\newcommand{\setattr}{\Delta^{A}}
\newcommand{\setattrvar}{\mathcal{X}^{A}}
\newcommand{\setval}{\Delta^{V}}
\newcommand{\setvalvar}{\mathcal{X}^{V}}
\newcommand{\REC}{\mathcal{REC}}

\newcommand{\graph}[1][G]{#1}
\newcommand{\nodes}[1][G]{N_{#1}}
\newcommand{\ports}[1][G]{P_{#1}}
\newcommand{\edges}[1][G]{E_{#1}}
\newcommand{\lablf}[1][G]{\mathcal{L}_{#1}}
\newcommand{\leftg}{\graph[L]}
\newcommand{\rightg}{\graph[R]}
\newcommand{\rwrule}{\leftg \Rightarrow \rightg}

\newcommand{\polGraph}{\graph}
\newcommand{\polNodes}[1][\polGraph]{\nodes[#1]}
\newcommand{\polEdges}[1][\polGraph]{\edges[#1]}
\newcommand{\polLe}{le}
\newcommand{\polLn}{ln}
\newcommand{\Pg}{\P_{\polGraph}}
\newcommand{\Cg}{\C_{\polGraph}}
\newcommand{\Ag}{\A_{\polGraph}}
\newcommand{\Rg}{\R_{\polGraph}}

\newcommand{\PCAg}{\PCA_{\polGraph}}
\newcommand{\ARCAg}{\ARCA_{\polGraph}}
\newcommand{\PARg}{\PAR_{\polGraph}}
\newcommand{\BARCAg}{\BARCA_{\polGraph}}
\newcommand{\BARg}{\BAR_{\polGraph}}
\newcommand{\UNDETg}{\UNDET_{\polGraph}}

\newcommand{\evsch}{\textit{G}}
\newcommand{\ev}{\textit{E}}
\newcommand{\actres}{P\hspace{-0.02in}\textsubscript{\textit{r}}}
\newcommand{\obl}{\textit{O}}
\newcommand{\dut}{\textit{D}}
\newcommand{\CBACOg}{\CBACO_{\polGraph}}
\newcommand{\OARCAg}{\OARCA_{\polGraph}}
\newcommand{\OPARg}{\OPAR_{\polGraph}}
\newcommand{\DPARg}{\DPAR_{\polGraph}}
\newcommand{\ETg}{\ET_{\polGraph}}
\newcommand{\EIg}{\EI_{\polGraph}}
\newcommand{\Eg}{\E_{\polGraph}}
\newcommand{\GEg}{\GE_{\polGraph}}
\newcommand{\Hg}{\H_{\polGraph}}

\newcommand{\act}{\textsf{act}}
\newcommand{\subj}{\textsf{subj}}
\newcommand{\obj}{\textsf{obj}}
\newcommand{\evtime}{\textsf{time}}

\newcommand{\locgraph}{G_{P}^{Q}}
\newcommand{\locgraphp}{G_{P'}^{\prime Q'}}
\newcommand{\locrwrule}{\leftg_{W} \Rightarrow \rightg_{M}^{N}}

\begin{document}

\title{A Graphical Framework for the Category-Based Metamodel for Access Control and Obligations\thanks{This work is financed by National Funds through the Portuguese funding agency, FCT - Fundação para a Ciência e a Tecnologia, within project UIDB/50014/2020.}}

\author{Sandra Alves\inst{1,2}
\and Jorge Igl\'esias\inst{2}}
    
\institute{CRACS - INESCTEC \\ Porto, Portugal\\
\email{sandra@fc.up.pt}\\\ \\
\and
DCC-FCUP \\ University of Porto, Porto, Portugal\\
   \email{up201204704@edu.fc.up.pt}
}
\maketitle

\begin{abstract}
  We design a graph-based framework for the visualisation and analysis of obligations in access control policies. We consider obligation policies in CBACO, the category-based access control model, which has been shown to subsume many of the most well known access control such as MAC, DAC, RBAC. CBACO is an extension of the CBAC metamodel that deals with obligations. We describe the implementation of the proposed model in PORGY, a strategy driven graph-rewriting tool, based on the theory of port-graphs. CBACO policies allow for dynamic behavior in the modelled systems, which is implemented using the strategy language of PORGY. 
  

\end{abstract}

\section{Introduction}
The ability to efficiently protect resources from unauthorised access and to preserve the integrity and confidentiality of critical data has never been more crucial, leading to an increasing interest in models to effectively control access to information and resources.
However, despite the variety of models,  there is still a pressing need for effective methods and tools to facilitate the tasks of policy specification and analysis. Formal specifications of access control models and policies have used theorem provers, purpose-built logics, functional approaches, etc, and although textual languages and logic-based models are convenient for theoreticians or computer experts, graphical models are more appealing to less technical users, therefore urging the development of models that establish a bridge between theoretical tools and the practitioners’ (mainly security administrators) needs. The motivation for developing a graph-based framework is clear: graphs are a natural model for distributed systems, with graphical languages being widely used for describing complex structures in a visual and intuitive way in a variety of domains (software modelling, representation of proofs, microprocessor design, XML documents, communication networks, social networks, biological systems, etc), while graph transformations, or graph rewriting, is used to define the dynamic behaviour of the system modelled. 

In this paper we develop a graph-based framework for the visualisation and analysis of obligations in access control policies.  Our basis will be CBACO~\cite{AlvesDF15}, an axiomatic metamodel for access control that deals with obligations in a comprehensive way and which extends the CBAC metamodel~\cite{Barker09} that subsumes the most well-known access control models. The notion of obligation helps to bridge a gap between requirements and policy enforcement. For example, consider a hospital scenario in which any doctor may be authorised to read the medical record of a patient in an emergency situation, but in that case there is a requirement to inform the patient afterwards.  Although access control models deal mainly with authorisations, incorporating the notion of an obligation facilitates the tasks of ensuring that obligations are enforced and ensuring that obligations are compatible with authorisations. CBAC(O) focuses on the notion of a \emph{category}, which is a class of entities that share some property,  and permissions are assigned to categories of users, rather than to individual users.  Categories can be defined on the basis of user attributes, geographical constraints, resource attributes, etc. In this way, permissions can change in an autonomous way unlike, e.g., role-based models, which require the intervention of a security administrator.  A graphical representation of CBAC policies, including composed policies involving several local policies in a multi-site system, was developed in~\cite{AlvesF17}, where static properties were checked using basic graph properties on paths. However, to deal with realistic scenarios one needs to be able to answer more complex and relevant questions, based on static and/or dynamic information. In particular, there is a need for tools to analyse the interactions between different policies which may deal with different features of the system. 

Taking the graphical model of CBAC as a basis, we develop a new graphical model for CBACO and implement it using PORGY~\cite{AndreiO:PORGY}, a strategy driven graph-rewriting tool based on the theory of port-graphs that provides a visual modelling environment with specific simulation and verification tools, which we believe to be well suited for the analysis of policies.  PORGY is based on the theory of port-graphs~\cite{Andrei08} and implements a strategy driven language for port graph rewriting. In a port-graph, edges attach to nodes at specific connection points called ports. Rules and rewriting strategies are also viewed as port-graphs that can be rewritten into new rules, enabling the modelling of adaptive systems. Port-graphs have been used to model complex systems in the areas of biochemistry and interaction nets, as well as algorithm animation and games. The main contributions of this work are:
\begin{itemize}
    \item We extend CBACO to deal with dynamic aspects of obligations;
    \item We present a new graphical model for CBACO;
    \item We implement our model using the PORGY framework, taking advantage of its strategy driven graph-rewriting language to implement the dynamic aspects of our graphical model.
\end{itemize}

\emph{Overview:}  In Section~\ref{sec:prelim}, we recall the \CBACO\ metamodel, extend it to deal with the state of duties and briefly describe PORGY. Section~\ref{sec:oblig} discusses a graphical model for obligations, while Section~\ref{sec:porgy} shows how our graphical model is implemented in PORGY. In Section~\ref{sec:dyno} we describe how the strategy language of PORGY can be used to implement the dynamics of the modeled systems. In Section~\ref{sec:rw}, we discuss related work, and in
 Section~\ref{sec:concl}, conclusions are drawn and further work is suggested.

\section{Preliminaries}
\label{sec:prelim}
We briefly describe below the key concepts underlying the metamodel for category-based access control with obligations and the formalism of port-graphs.
\subsection{CBACO: Obligations in the category-based metamodel}
The \CBACO\ metamodel~\cite{AlvesDF15} is based on the notion of category. Informally, a category is any of several distinct classes or groups to which entities may be assigned.  Entities are denoted by constants in
a many sorted domain of discourse, including: a countable set
$\mathcal C$ of categories, denoted $c_0$, $c_1$, $\dots$, a countable
set $\mathcal P$ of principals, denoted $p_0$, $p_1$, $\dots$, a countable set $\mathcal A$ of named
\emph{actions}, denoted $a_0$, $a_1$, $\dots$, a countable set
$\mathcal R$ of \emph{resource identifiers}, denoted $r_0$, $r_1$,
$\dots$, a finite set $\mathcal Auth$ of possible \emph{answers} to
access requests (e.g., \{\grant, \deny, {\sf undetermined}\}), a finite set  $\E$  of events (denoted $e_{1}, e_{2}, \dots$), a finite set $\GE$ of event schemes  (denoted $ge_{1}, ge_{2}, \dots$) and a finite set $\H$ of event histories (denoted $h_{1}, h_{2}, \dots$). More generally, entities can be represented by
a data structure (e.g., a principal could be represented by a term
$principal(p_i, attributeList)$), but constants will be sufficient for
most examples in this paper.
We start by defining some notions related to events, permissions, prohibitions, authorisations and obligations. The definitions in this subsection follow the notions defined in~\cite{AlvesDF15}.

A (specific) \textbf{event} $e \in \E$ is a ground event specification $\gspec$, which represents an actual action/happening that occurred in a system. 

A \textbf{generic event} $ge \in \GE$ represents a family of events and is given by an event specification $\spec$\footnote{The definition of what an event specification is, depends on the particular system.}, to which is associated a context $ge[X_{1}, \dots, X_{n}]$ where $X_{1}, \dots, X_{n}$ are the term variables in $\spec$. Replacing $X_{1}, \dots, X_{n}$ with ground values $\gv_{1}, \dots, \gv_{n}$ will result in a ground event specification $\gspec$.

An \textbf{event history} $h \in \H$ is a possible sequence of events of the system represented as a list of events of the form $ [e_{1}, \dots, e_{n}] $ where every $e_{i}$ is associated to a specific time $t_i$ and such that if $i < j$ then $t_{i}\preceq t_{j}$, according to some notion of ordering on time. A subsequence of $h$ is an event interval, denoted $I = (e_{i}, e_{j}, h)$, where $e_{i}$ is the first event of the interval and $e_j$ is the last one. We say that $e_i$ opens the interval and $e_j$ closes it. For every event $e_k$ in $h$ such that $t_{i}\preceq t_{k}\preceq t_{j}$ we say that $e_k \in (e_{i}, e_{j}, h)$.

A \textbf{permission} is a pair $(a,r)$ of an action and a resource, and an
\textbf{authorisation} is a triple $(p,a,r)$ that associates a permission
with a principal.

A \textbf{generic obligation} $o \in \O$ is an action $a \in \A$ on a resource $r \in \R$ that must be performed between two event schemes $ge_{1}, ge_{2} \in \GE$, represented as a tuple $o = (a, r, ge_{1}, ge_{2})$. If there is no starting (resp. closing) event scheme, meaning the obligation can be performed at any time before $ge_{2}$ (resp. after $ge_1$), then $ge_{1}$ (resp. $ge_{2}$) is $\bot$. A specific/concrete obligation is an obligation where the event schemes $ge_{1}, ge_{2}$ are ground terms, therefore corresponding to specific events.
Like authorizations, obligations will be assigned to categories, but these might not be the same as authorization categories.

A \textbf{duty} $d \in \D$, is a tuple $(p, o)$, where $p$ is a principal, $o$ is a concrete obligation $(a, r, e_{1}, e_{2})$ such that there exists a generic obligation $(a, r, ge_{1}, ge_{2})$ assigned to $p$ and $e_{1}, e_{2}$ are instantiations of event schemes $ge_{1}, ge_{2}$, respectively. If there is no starting (resp. closing) event then $e_{1}$ (resp. $e_{2}$) is $\bot$.

 The metamodel includes the following relations:
\begin{itemize}
\item  $\mathcal P \mathcal C \mathcal
  A$ $\subseteq \mathcal P \times \mathcal C$, such that $(p,c) \in
  \mathcal P \mathcal C \mathcal A$ iff a principal $p \in \mathcal P$
  is assigned to the category $c \in \mathcal C$.
\item $\mathcal A \mathcal R \mathcal C \mathcal A$
  $\subseteq \mathcal A \times \mathcal R \times \mathcal C$, such
  that $(a,r,c) \in \mathcal A \mathcal R \mathcal C \mathcal A$ iff
  action $a \in \mathcal A$ on resource $r \in \mathcal R$ can be
 performed by the principals assigned to the category $c \in \mathcal C$.
\item $\mathcal P \mathcal A \mathcal R$ $\subseteq
  \mathcal P \times \mathcal A \times \mathcal R$, such that $(p,a,r)
  \in \mathcal P \mathcal A \mathcal R$ iff a principal $p \in
  \mathcal P$ can perform the action $a \in \mathcal A$ on the
  resource $r \in \mathcal R$.
\item $ \BARCA \subseteq \A \times \R \times \C $ is the prohibition-category assignment: $ (a, r, c) \in \BARCA $ iff members of category $ c $ cannot perform action $ a $ on resource $ r $;
\item $ \BAR \subseteq \P \times \A \times \R $ is the prohibition-principal assignment: $ (p, a, r) \in \BAR $ iff principal $ p $ cannot perform action $ a $ on resource $ r $;
\item $ \UNDET \subseteq \P \times \A \times \R $ is the undetermined permission principal assignment: $ (p, a, r) \in \UNDET $ iff principal $ p $ is neither authorized nor banned from performing action $ a $ on resource $ r $;

\item $\OARCA \subseteq \O \times \C$ is the obligation-category assignment: $(o, c) \in \OARCA$ iff there exists an obligation $o = (a, r, ge_1 , ge_2 )$ such that principals of category $c$ are obliged to perform action $a$ on resource $r$ between events that are instantiations of event schemes $ge_{1}$ and $ge_{2}$; 
\item $\OPAR \subseteq \P \times \O $ is the obligation-principal assignment: $(p, o) \in \OPAR$ iff there exists an obligation $o = (a, r, ge_1 , ge_2 )$ such that principal $p$ is obliged to perform action $a$ on resource $r$ between events that are instantiations of event schemes $ge_{1}$ and $ge_{2}$;
\item $\DPAR \subseteq \D$ is the duty assignment: $d = (p, o) \in \DPAR$ iff there is a concrete obligation $o = (a, r, e_1 , e_2 )$ such that principal $p$ is obliged to perform action $a$ on resource $r$ between events $e_{1}$ and $e_{2}$; 
\item $\ET \subseteq \E \times \GE$ is the event instantiation relation: $(e, ge) \in \ET$ iff event $e$ is an instance of event scheme $ge$, denoted $ e :: ge$, according to an instantiation relation between events and event schemes;
\item $\EI \subseteq \E \times \E \times \H$ is the event interval relation: $(e_{1}, e_{2}, h) \in \EI$ iff event $e_{2}$ closes the interval initiated by event $e_{1}$ in event history $h$.
\end{itemize}

The relations defined above satisfy the
following axioms, where we assume that there exists a partial ordering $\subseteq$ between authorisation categories and  a partial ordering $\oblOr$ between obligation categories:
\begin{multline} \label{ax:par}
  \forall p \in \mathcal P,~\forall a \in \mathcal A,~\forall r \in
\mathcal R, \hfill (\exists c, c' \in \mathcal C, ((p,c)\in\mathcal{PCA} \wedge ~ c \subseteq c' ~\wedge  (a,r,c')\in \mathcal{ARCA})\\ \hfill
 \Leftrightarrow (p,a,r)\in\mathcal{PAR})
\end{multline}
\vspace{-0.4in}
\begin{multline} \label{ax:bar}
\forall p \in \P, \forall a \in \A, \forall r \in \R, ((\exists c, c' \in \C, (p, c) \in \PCA \wedge c' \subseteq c \wedge (a, r, c') \in \BARCA) \\ \Leftrightarrow (p, a, r) \in \BAR)
\end{multline}
\vspace{-0.4in}
\begin{multline}
\forall p \in \P, \forall a \in \A, \forall r \in \R, (((p, a, r) \notin \PAR \wedge (p, a, r) \notin \BAR) \\ \Leftrightarrow (p, a, r) \in \UNDET)
\end{multline}
\vspace{-0.2in}
\begin{equation}
\PAR \cap \BAR = \emptyset
\end{equation}
\vspace{-0.2in}
\begin{multline} \label{ax:obl}
\forall o \in \mathcal{O}  ( ( \exists c, c' \in \C,
(p,c)\in\PCA \wedge c \oblOr c' \wedge
(o,c')\in \OARCA ) \Leftrightarrow (p,o)\in\OPAR )
\end{multline}
\vspace{-0.4in}
\begin{multline}
\forall p \in \P, \forall a \in \A, \forall r \in \R, \forall e_1,e_2 \in \mathcal{E},  ( ( \exists ge_1, ge_2 \in\GE,(p,\overbrace{(a,r,ge_1,ge_2)}^o)\in\OPAR, \\ e_1::ge_1, e_2:: ge_2
) \Leftrightarrow \underbrace{(p,(a,r,e_1,e_2)}_d)\in\DPAR )
\end{multline}
\subsection{State of Duties}
Issued obligations, can be in one of four states: invalid, fulfilled, pending or violated. This is unlike permissions, which are either issued or not.
To deal with that, we defined the following new relations to deal with the state of duties:
\begin{itemize}
	\item Fulfilled: $\mathcal{FULFILLED} \subseteq \D \times \E \times \H$, such that $(d,e,h)\in \mathcal{FULFILLED}$ if the duty $d$ was fulfilled by the event $e$ in history $h$.
	
	\item Pending: $\mathcal{PENDING} \subseteq \D \times \H$, such that $(d,h)\in \mathcal{PENDING}$ if the duty $d$ is pending in history $h$.
	
	\item Violated: $\mathcal{VIOLATED} \subseteq \D \times \H$, such that $(d,h)\in \mathcal{VIOLATED}$ if the duty $d$ is violated in history $h$.
\end{itemize}
We do not consider a relation on invalid duties, since our model does not allow them to be issued. Furthermore, we introduce the following axioms regarding the state of duties, where we assume that every history of events starts with an event $\bot$.
\[
\begin{array}{l}
\forall \overbrace{(p,a,r,e_1,e_2)}^d \in \DPAR, \forall h \in \H ( (  e_1\in h,\\
\ \ \ \ \ \ \ \ \ \ \ \ \ \exists e_3 \in h, (e_1,e_3,h)\in\EI,((e_3,e_2,h)\in\EI\vee(e_1,e_2,h)\notin\EI),\\
\ \ \ \ \ \ \ \ \ \ \ \ \ \ \ \ \ \ \ e_3.\subj=p, e_3.\act=a, e_3.\obj=r
) \Leftrightarrow (d, e_3, h) \in \mathcal{FULFILLED} )\\
\forall \overbrace{(p,a,r,e_1,e_2)}^d \in \DPAR, \forall h \in \H ( ( e_1\in h, (e_1, e_2, h) \notin \EI,
\nexists e_3 \in h,\\ \ \ \ \ \ \ (e_1, e_3, h) \in \EI, e_3.\subj=p, e_3.\act=a, e_3.\obj=r
) \Leftrightarrow (d, h) \in \mathcal{PENDING} )
\\
\forall \overbrace{(p,a,r,e_1,e_2)}^d \in \DPAR, \forall h \in \H ( ( (e_1, e_2, h)\in\EI,
\nexists e_3 \in (e_1, e_2, h),\\ \ \ \ \ \ \ \ \ \ \ \ \ \ \ \ \ \ \ \ \ \ \ \ \ \ \ e_3.\subj=p, e_3.\act=a, e_3.\obj=r
) \Leftrightarrow (d, h) \in \mathcal{VIOLATED} )
\end{array}
\]
\subsection{PORGY}
PORGY~\cite{AndreiO:PORGY} is an interactive tool that has been developed to visualise and analyse port graph reduction systems. 
Visually, a port graph is a graph where edges are attached to
nodes at points called \emph{ports}.  Below we give a brief introduction to port graphs (see~\cite{Andrei08,AndreiK08c}
for more details and examples).
We assume a  signature $ \sign = \langle \setattr, \setattrvar, \setval, \setvalvar \rangle $ where $\setattr $ is a set of attributes; $\setattrvar$  is a set of attribute variables; $\setval$  is a set of values; and $\setvalvar$ is a set of value variables.

We can represent elements of the graph as records (sets of pairs $(\textit{attribute},\textit{value})$) over the signature $\sign$. We assume the attributes \textit{Name}, \textit{Interface}, \textit{Arity}, \textit{Attach} and \textit{Connect}. Attribute \textit{Name} identifies records in the sense that if two records have the same value for attribute \textit{Name} then they have the same set of defined attributes, although their values can differ.

A \textbf{port graph} over a signature $ \Delta $ is defined as a tuple $ \graph = \langle \nodes, \ports, \edges, \lablf \rangle $ where: $\nodes$ is a finite set of nodes (denoted $ n_{1}, n_{2}, \dots $); $\ports$ is a finite set of ports (denoted $ p_{1}, p_{2}, \dots $); $\edges$ is a finite set of undirected edges (denoted $ e_{1}, e_{2}, \dots $); $\lablf$ is a labeling function. For every element of the sets described above, $\lablf$ returns the record associated with it such that: for every node $ n \in \nodes $ the record contains attribute \textit{Interface} and $ \lablf(n).\textit{Interface} $ returns the set of ports of $ n $; for every port $ p \in \ports $ the record contains attributes \textit{Attach} and \textit{Arity}, such that $ \lablf(p).\textit{Arity} $ gives the number of edges connected to that port and $ \lablf(p).\textit{Attach} $ returns the node $ n $ to which $ p $ belongs; for every edge $ e \in \edges $ the record contains attribute \textit{Connect} and $ \lablf(e).\textit{Connect} $ returns $ \{p_{1},p_{2}\} $ the ports that $ e $ is connected to.

A \textbf{port graph rewrite rule} is a port graph composed of two subgraphs, $ \leftg $ and $ \rightg $, and a special node (called arrow node) that connects the two. It defines a way of rewriting graphs in the sense that, if there is a instance of $ \leftg $, it can be replaced by $ \rightg $ and the arrow node defines the way to reconnect $ \rightg $ to the rest of the graph. Ports in the arrow node have attribute Type that can have values \textsf{bridge}, \textsf{blackhole} or \textsf{wire}. These values define the rewriting in the following way:
\begin{itemize}
\item a port with Type \textsf{bridge} has exactly one edge connecting it to a port $ p $ of $ \leftg $ and one or more connecting it to ports $ p_{i} $ of $ \rightg $. This type of port defines that edges of the graph being rewritten that are connected to the image of $ p $ and are not in $ \leftg $ should be reconnected to every $ p_{i} $;
\item there is at most one port with Type \textsf{blackhole} and it only has edges connected to ports of $ \leftg $. It defines that edges connected to such ports should be erased;
\item a port with Type \textsf{wire} has exactly two edges connecting it to two ports $ p_{1}, p_{2} $ of $ \leftg $ and it defines that there should be created an edge between every port $ p_{1i} $ that is connected to the image of $ p_{1} $ and every port $ p_{2i} $ that is connected to the image of $ p_{2} $.
\end{itemize}
A rewrite rule $ \rwrule $ matches a graph $ \graph $ if there is a morphism $ f $ such that $ f(\leftg) \subseteq \graph $ and the image of every port $ p $ in $ \leftg $ not connected to the arrow node is not connected to any port of $ \graph \setminus f(\leftg) $. This last condition ensures that rewriting does not leave dangling edges.

Given  a port graph $ \graph $, a rewrite rule $ \rwrule $ and a morphism $ f $ such that $ f(\leftg) \subseteq \graph $, a rewriting step from $ \graph $ by rule $ \rwrule $ and morphism $ f $, denoted $ \graph \rightarrow_{\rwrule}^{f} \graph' $, is defined as follows: the \textit{build} phase: graph $ f(\rightg) $ is added to $ \graph $; the \textit{rewiring} phase: edges connecting $ \graph \setminus f(\leftg) $ to $ f(\leftg) $ are connected to $ f(\rightg) $ following the rules set by the \textit{arrow node}; the \textit{deletion} phase: $ f(\leftg) $ is erased obtaining $ \graph' $.

\begin{figure}[ht]
\vspace{-0.2in}
  \begin{subfigure}[b]{0.3\linewidth}
    \includegraphics[width=\linewidth]{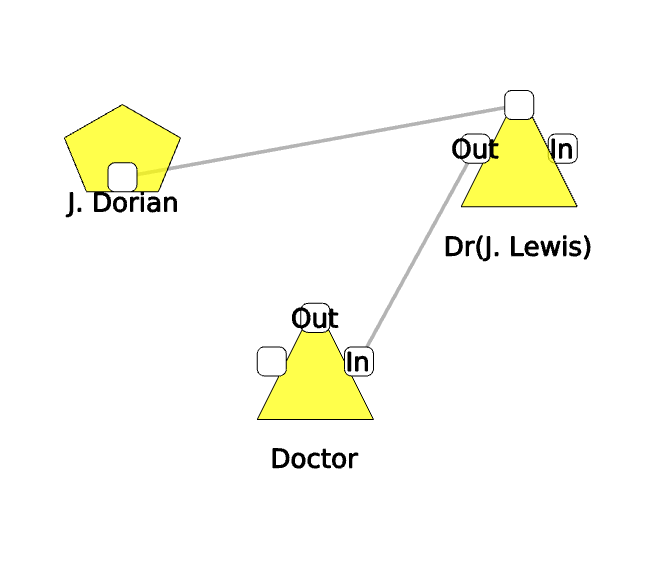}
  \end{subfigure}
  \hfill
  \begin{subfigure}[b]{0.3\linewidth}
    \includegraphics[width=\linewidth]{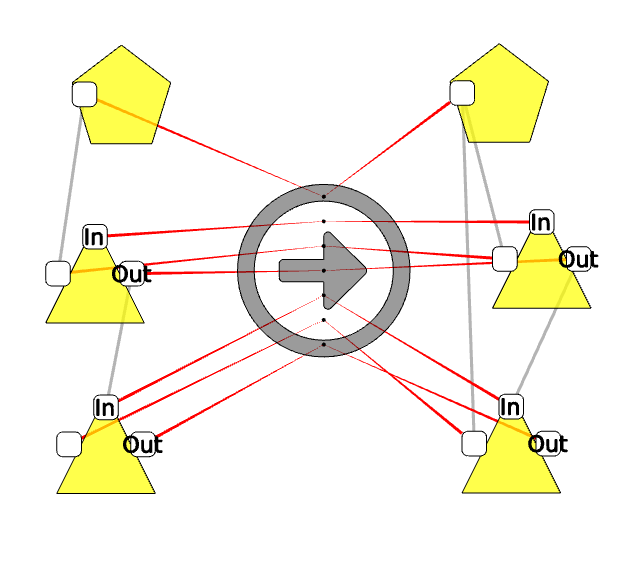}
  \end{subfigure}
    \hfill
  \begin{subfigure}[b]{0.3\linewidth}
    \includegraphics[width=\linewidth]{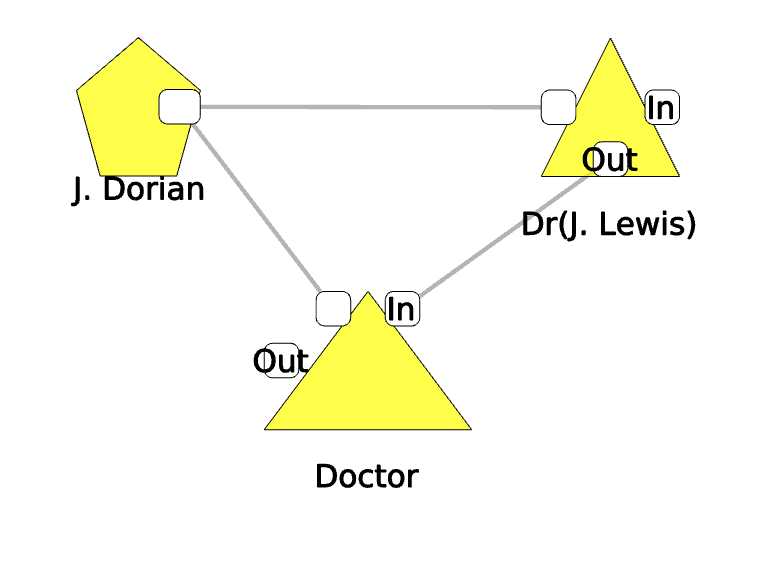}
  \end{subfigure}
  \caption{Port Graph Rewriting}
  \label{fig:ex1}
\end{figure}

\begin{example}
  In Figure~\ref{fig:ex1} the port-graph on the right is obtained from the port-graph on the left, by applying the port-graph rewrite rule represented by the port-graph in the middle\footnote{In this rule the top two nodes of \leftg\ belong to $W$, the top two nodes of \rightg\ belong to $M$ and the bottom one to $N$.}.  
\end{example}
\vspace{-0.1in}
Given a set of rewrite rules and a graph, it will be possible to apply the rules in different orders and places of the graph. To control that, PORGY uses a strategy language. 
 To define these strategies we start by defining the notions of position and banned subgraphs of a located graph.

A \textbf{located graph} $ \locgraph $ is a graph such that $ P \subseteq G $ is called the position subgraph and $ Q \subseteq G $ is the banned subgraph.
This notion is used in PORGY to select nodes, ports or edges where rewriting should happen and where it cannot happen.

A \textbf{located rewrite rule} $ \locrwrule $ can be applied to a located graph $ \locgraph $ by means of a morphism $ f $ if: $ f(L) \cap Q = \emptyset $; $ f(L) \cap P = W $, if $ W $ is specified. In this case we write $ \locgraph \rightarrow_{\locrwrule}^{f} \locgraphp $ where $ P' = (P \setminus f(\leftg)) \cup f(M) $ and $ Q' = Q \cup f(N) $.

In lay terms $ W $ defines the elements of $ L $ which images should be in the position graph and $ M $ (resp. $ N $) defines the elements of $ R $ which images are to be added to the position (resp. banned) subgraph of $ G' $. 
In PORGY we can use the strategy language to control both the way rewrite rules are applied and the position and banned subgraphs (see~\cite{AndreiO:PORGY} for a complete description of PORGY's strategy language).

\section{Graph Representation of Obligations}
\label{sec:oblig}
We now define a graphical model for $\CBACO$ that extends the $\CBAC$ representation using graphs presented in~\cite{AlvesF17}. 
\begin{definition}[$\CBACO$ policy graph] \label{def:polgraphcbaco}
  A $\CBACO$ policy graph $ \polGraph $ is a tuple $ \langle \polNodes, \polEdges,\polLn, \polLe \rangle $ with:
  \begin{itemize}
  \item $ \polNodes $ a finite set of nodes (denoted $ n_{1}, n_{2}, \dots $);
  \item $ \polEdges $ a finite set of undirected edges (denoted $ e_{1}, e_{2}, \dots $);
  \item $ \polLn: \polNodes \rightarrow \REC $ a labeling function for nodes such that for every node $ n \in \polNodes $, $ \polLn(n).ent \in \P \cup \C \cup \A \cup \R \cup \GE \cup \E \cup ( \A \times \R ) \cup ( \A \times \R \times \GE \times \GE ) \cup ( \P \times \A \times \R \times \E \times \E)$ and $ \polLn(n).type \in \{P, C, A, R, \evsch, \ev, \actres, \obl, \dut \} $ where:
    \begin{itemize}
    \item $ \polLn(n).type = P $ iff $ \polLn(n).ent = p \in \P $;
    \item $ \polLn(n).type = C $ iff $ \polLn(n).ent = c \in \C $;
    \item $ \polLn(n).type = A $ iff $ \polLn(n).ent = a \in \A $;
    \item $ \polLn(n).type = R $ iff $ \polLn(n).ent = r \in \R $;
    \item $ \polLn(n).type = \evsch $ iff $ \polLn(n).ent = ge \in \GE $;
    \item $ \polLn(n).type = \ev $ iff $ \polLn(n).ent = e \in \E $;
    \item $ \polLn(n).type = \actres $ iff $ \polLn(n).ent = (a, r) \in \A \times \R $;
    \item $ \polLn(n).type = \obl $ iff $ \polLn(n).ent = (a, r, ge_{1}, ge_{2}) \in \A \times \R \times \GE \times \GE $;
    \item $ \polLn(n).type = \dut $ iff $ \polLn(n).ent = (p, a, r, e_{1}, e_{2}) \in \P \times \A \times \R \times \E \times \E $;
    \end{itemize}
  \item $ \polLe: \polEdges \rightarrow \REC $ a labeling function for edges such that for every edge $ e \in \polEdges $, $ \polLe(e).adj = \{n_{1}, n_{2}\} $ gives the two nodes connected by the edge. We also define the function $type$ such that $type(e) = \{\polLn(n_{1}).type, \polLn(n_{2}).type\}$. Furthermore if:
    \begin{itemize}
    \item $ type(e) = CC $ then $\polLe(e).\textit{target} \subseteq \{\polLn(n_{1}).ent, \polLn(n_{2}).ent\} $ and\\ $\polLe(e).auth, \polLe(e).obl \in \{\top, \bot\}$ ;
    \item $ type(e) = C\actres $ then $\polLe(e).auth \in \{A, B\}$;
    \item $ type(e) = \obl\evsch $ then $\polLe(e).ge \in \{i, f\}$;
    \item $ type(e) = \dut\ev $ then $\polLe(e).ev \in \{i, f\}$;
    \item $ type(e) = \evsch\evsch $ then $ \polLe(e).\textit{target} \in \{\polLn(n_{1}).ent, \polLn(n_{2}).ent\}$;
    \item $ type(e) = \ev\ev $ then $\polLe(e).\textit{target} \in \{\polLn(n_{1}).ent, \polLn(n_{2}).ent\}$;
    \end{itemize}
  \end{itemize}
\end{definition}

Nodes of type $\ev$ have an attribute $now$ set to either $ \top $ or $ \bot $. Only one node of type $\ev$ can have $now$ set to $\top$, which specifies what is the current event. 		
Nodes of type $\obl$ represent obligations and are always connected to a node of type $\actres$ and zero, one or two nodes of type $\evsch$. Other connections will be to nodes of type $C$. Nodes of type $\dut$ represent duties and are always connected to a node of type $P$ a node of type $\actres$ and zero, one or two nodes of type $E$.

Because we represent both $\subseteq$ and $\oblOr$ there is the need to distinguish edges of type $CC$. Edges $e$ that represent the $\subseteq$ (resp. $\oblOr$) will have the $auth$ (resp. $obl$) attribute value set to $\top$ and we write $type(e) = CC_{\actres}$ (resp. $type(e) = CC_{\obl}$). \\
Edges $e$ of type $\textit{C\actres}$ serve the same purpose of the edges of type $\textit{CA}$ in the graph representation of \CBAC\ in~\cite{AlvesF17} and for that reason have attribute $auth$ defined as either $A$ or $B$ and we will write $ type(e) = \textit{C\actres}^A $ and $type(e) = \textit{C\actres}^B $, respectively, to distinguish edges corresponding to authorizations/prohibitions.

Edges of type $\obl\evsch$ (resp. $\dut\ev$) have the $ge$ (resp. $ev$) attribute value set to either $i$ or $f$ to distinguish between the initial and final event scheme (resp. event). 
Edges of type $\evsch\evsch$ are used to represent the notion that some event schemes are instantiations of other event schemes and have attribute $\textit{target}$ set to the node that represents the event scheme that is more general (the one which the other is an instantiation of). Furthermore, the type of this edges when in terms of types of paths will have an arrow on top in the same way edges of type $CC$ have. Edges of type $\ev\ev$ connect events of the same history and have attribute $\textit{target}$ set to the node that represents the one event that comes after the other. As edges of type $CC$ and $\evsch\evsch$ the type of this edges in paths will have an arrow on top.

\begin{definition}[Path] \label{def:path}
Given a policy graph $\polGraph =  \langle \polNodes, \polEdges, \polLn, \polLe \rangle $ a path of length $d$ is a sequence of pairwise distinct nodes $n_{0}, n_{1}, \dots, n_{d}$ such that for every $ 1 \leq i \leq d $, $\polLe(e).adj=\{n_{i-1}, n_{i}\}$ for some $e \in \polEdges$.
\end{definition}

\begin{definition}[Constrained path and constrained reverse path] \label{def:constpath}
A constrained path of length $d$ is a sequence $n_{0}, e_{1}, n_{1}, \dots, e_{d}, n_{d}$, with pairwise distinct nodes and edges such that for every $1 \leq i \leq d$, $\polLe(e_{i})=\{n_{i-1}, n_{i}\}$ and $n_{i} \in \polLe(e_{i}).\textit{target}$ when $\polLe(e_{i}).\textit{target}$ exists.\\
A constrained inverse path of length $d$ is a sequence $n_{0}, e_{1}, n_{1}, \dots, e_{d}, n_{d}$ with pairwise distinct nodes and edges such that for every $1 \leq i \leq d$, $\polLe(e_{i})=\{n_{i-1}, n_{i}\}$ and $n_{i-1} \in \polLe(e_{i}).\textit{target}$ when $\polLe(e_{i}).\textit{target}$ exists.
\end{definition}
\begin{definition}[Types of paths] \label{def:pathtypes}
Consider a path $n_{0}, n_{1}, \dots, n_{d}$ of length $d$ such that $\polLn(n_{i}).type = T_{i}$ for every $0 \leq i \leq d$. The type of this path is given by the types of its edges $T_{0}T_{1}, T_{1}T_{2}, \dots, T_{d-1}T_{d}$. \\
The notation $type(n_{0}, n_{1}, \dots, n_{d}) = T_{0}T_{1}, T_{1}T_{2}, \dots, T_{d-1}T_{d}$ is used to denote that there is a path $ n_{0}, n_{1}, \dots, n_{d} $ and its edges are of type $T_{0}T_{1}, T_{1}T_{2}, \dots, T_{d-1}T_{d}$.\\
If an edge $e_{i}$ of the path is connecting two nodes $n_{i-1}$ and $n_{i}$ both of type $C$ then its type will be $\overrightarrow{CC}$ if $n_{i} \in \polLe(e_{i}).\textit{target}$ and $\overleftarrow{CC}$ if $n_{i-1} \in \polLe(e_{i}).\textit{target}$. 
\end{definition}
Note that an edge might be both of type $\overrightarrow{CC}$ and $\overleftarrow{CC}$.

\begin{definition}[Redundant edges in the $\CBACO$\ policy graph] \label{def:rededgescbaco}
Consider a $\CBACO$ policy graph $ \polGraph = \langle \polNodes, \polEdges, \polLn, \polLe \rangle$. An edge $ e \in \polEdges$ is redundant if:
\begin{itemize}
	\item $ type(e) = PC$, i.e. it connects nodes $ n_{1}, n_{2} $ such that $ \polLn(n_{1}).ent = p$ and $ \polLn(n_{2}).ent = c$ with $p \in \P$, $c \in \C$, and there is a path of type $PC, (\overrightarrow{CC})^* $ and length greater or equal to 2 connecting $ n_1 $ and $ n_2$;
	\item $ type(e) = \overrightarrow{CC}_\actres$, i.e. it connects nodes $ n_{1}, n_{2} $ such that $ \polLn(n_{1}).ent = c_1$ and $ \polLn(n_{2}).ent = c_2$ with $c_{1}, c_{2} \in \C$ and $\polLe(e).auth = \top$, and there is a path of type $ (\overrightarrow{CC}_{\actres})^* $ and length greater or equal to 2 connecting $ n_1 $ and $ n_2$;
	\item $ type(e) = \overrightarrow{CC}_\obl$, i.e. it connects nodes $ n_{1}, n_{2} $ such that $ \polLn(n_{1}).ent = c_1$ and $ \polLn(n_{2}).ent = c_2$ with $c_{1}, c_{2} \in \C$ and $\polLe(e).obl = \top $, and there is a path of type $(\overrightarrow{CC}_{\obl})^* $ and length greater or equal to 2 connecting $ n_1 $ and $ n_2$;
	\item $ type(e) = \textit{C\actres}^A$, i.e. it connects nodes $ n_{1}, n_{2} $ such that $ \polLn(n_{1}).ent = c$ and $ \polLn(n_{2}).ent = (a, r)$ with $c \in \C$, $(a, r) \in \A \times \R$ and $\polLe(e).auth = A$, and there is a path of type $ (\overrightarrow{CC}_{\actres})^{*}, \textit{C\actres}^A$ and length greater or equal to 2 connecting $ n_1 $ and $ n_2$;
	\item $ type(e) = \textit{C\actres}^B$, i.e. it connects nodes $ n_{1}, n_{2} $ such that $ \polLn(n_{1}).ent = c$ and $ \polLn(n_{2}).ent = (a, r)$ with $c \in \C$, $(a, r) \in \A \times \R $ and $\polLe(e).auth = B$, and there is a path of type $ (\overleftarrow{CC}_{\actres})^{*}, \textit{C\actres}^B $ and length greater or equal to 2 connecting $ n_1 $ and $ n_2$;
	\item $ type(e) = \textit{C\obl}$, i.e. it connects nodes $ n_{1}, n_{2} $ such that $ \polLn(n_{1}).ent = c$ and $ \polLn(n_{2}).ent = (a, r, ge_{1}, ge_{2})$ with $c \in \C$, $(a, r, ge_{1}, ge_{2}) \in \A \times \R \times \GE \times \GE$, and there is a path of type $ (\overrightarrow{CC}_{\obl})^{*}, \textit{C\obl}$ and length greater or equal to 2 connecting $ n_1 $ and $ n_2$;
	\item $ type(e) = \ev\evsch$, i.e. it connects nodes $ n_{1}, n_{2} $ such that $ \polLn(n_{1}).ent = e$ and $ \polLn(n_{2}).ent = ge$ with $e \in \E$, $ge \in \GE$, and there is a path of type $ \ev\evsch, (\overrightarrow{\evsch\evsch})^* $ and length greater or equal to 2 connecting $ n_1 $ and $ n_2$;
	\item $ type(e) = \overrightarrow{\evsch\evsch}$, i.e. it connects nodes $ n_{1}, n_{2} $ such that $ \polLn(n_{1}).ent = ge_1$ and $ \polLn(n_{2}).ent = ge_2$ with $ge_{1}, ge_{2} \in \GE$ and $ge_2\in\polLe(e).\textit{target}$, and there is a path of type $ (\overrightarrow{\evsch\evsch})^* $ and length greater or equal to 2 connecting $ n_1 $ and $ n_2$.
\end{itemize} 
\end{definition}

\begin{definition}[Well-formed $\CBACO$ policy graph]
A $\CBACO$ policy graph $\polGraph$ is well-formed iff for every $n_{1}, n_{2} \in \polNodes$ if $\polLn(n_{1}).ent=\polLn(n_{2}).ent$ and $\polLn(n_{1}).type=\polLn(n_{2}).type$ then $n_{1}=n_{2}$, and for every $e_{1}, e_{2} \in \polEdges$ if $\polLe(e_{1}).adj=\polLe(e_{2}).adj$ then  either $e_{1}=e_{2}$ or $e_{1}, e_{2}$ have type $\textit{C\actres}$ and $\polLe(e_{1}).auth \ne \polLe(e_{2}).auth$, for every $e \in \polEdges$ with $\polLe(e).adj=\{n_{1}, n_{2}\}$ one of the following is true:
\begin{itemize}
	\item $\polLn(n_{1}).type=P \wedge \polLn(n_{2}).type=C$;
	\item $\polLn(n_{1}).type=C \wedge \polLn(n_{2}).type=C \wedge  \polLe(e).\textit{target} \subseteq \{\polLn(n_{1}).ent, \polLn(n_{2}).ent\} \wedge \polLe(e).auth, \polLe(e).obl \in \{\top, \bot \} $;
	\item $\polLn(n_{1}).type=C \wedge \polLn(n_{2}).type=\actres \wedge \polLe(e).auth \in \{A, B\}$;
	\item $\polLn(n_{1}).type=C \wedge \polLn(n_{2}).type=\obl$;
	\item $\polLn(n_{1}).type=\actres \wedge \polLn(n_{2}).type=A \wedge \polLn(n_{1}).ent = (a, r) \wedge \polLn(n_{2}).ent = a$;
	\item $\polLn(n_{1}).type=\actres \wedge \polLn(n_{2}).type=R \wedge \polLn(n_{1}).ent = (a, r) \wedge \polLn(n_{2}).ent=r$;
	\item $\polLn(n_{1}).type=\obl \wedge \polLn(n_{2}).type=\actres \wedge \polLn(n_{2}).ent = (a, r) \wedge \polLn(n_{1}).ent = (a, r, ge_{1}, ge_{2})$;
	\item $\polLn(n_1).type=\obl \wedge \polLn(n_{2}).type=\evsch \wedge \polLn(n_{1})=(a, r, ge_{1}, ge_{2}) \wedge ((\polLe(e).ge=i \wedge \polLn(n_{2}).ent=ge_{1}) \vee (\polLe(e).ge=f \wedge \polLn(n_{2}).ent=ge_{2})) $;
	\item $\polLn(n_{1}).type=\dut \wedge \polLn(n_{2}).type=\P \wedge \polLn(n_{2}).ent = p \wedge \polLn(n_{1}).ent = (p, a, r, e_{1}, e_{2})$;
	\item $\polLn(n_{1}).type=\dut \wedge \polLn(n_{2}).type=\actres \wedge \polLn(n_{2}).ent = (a, r) \wedge \polLn(n_{1}).ent = (p, a, r, e_{1}, e_{2})$;
	\item $\polLn(n_1).type=\dut \wedge \polLn(n_{2}).type=\ev \wedge \polLn(n_{1})=(p, a, r, e_{1}, e_{2}) \wedge ((\polLe(e).e=i \wedge \polLn(n_{2}).ent=e_{1}) \vee (\polLe(e).e=f \wedge \polLn(n_{2}).ent=e_{2})) $;
	\item $\polLn(n_{1}).type=\ev \wedge \polLn(n_{2}).type=\ev \wedge \polLe(e).\textit{target} \in \{\polLn(n_{1}).ent, \polLn(n_{2}).ent\} $;	
	\item $\polLn(n_{1}).type=\ev \wedge \polLn(n_{2}).type=P$;
	\item $\polLn(n_{1}).type=\ev \wedge \polLn(n_{2}).type=A$;
	\item $\polLn(n_{1}).type=\ev \wedge \polLn(n_{2}).type=R$;
	\item $\polLn(n_1 ).type=\ev \wedge \polLn(n_2 ).type=\evsch$;
	\item $\polLn(n_{1}).type=\evsch \wedge \polLn(n_{2}).type=\evsch \wedge \polLe(e).\textit{target} \in \{\polLn(n_{1}).ent, \polLn(n_{2}).ent\} $;
\end{itemize}
and there are no redundant edges. Moreover if a constrained path and an inverse constrained path begin in the same node with type $P$ and end in nodes of type $\actres$ such that the last edges of the paths are of type $\textit{C\actres}^{A}$ and $\textit{C\actres}^{B}$, respectively, then the end node must be different. This means that a principal cannot be both able and unable to perform an action on a resource.
\end{definition}
A graph $\polGraph$ defines a particular $\CBACO$ policy.
\begin{definition}[$\CBACOg$] \label{def:cbacog}
Given a well-formed policy graph $\polGraph = \langle \polNodes, \polEdges, \polLn, \polLe \rangle $ the $\CBACO$ policy defined by $\polGraph$ is defined by the following entities and relations:
\begin{itemize} 
	\item $\Pg = \{\polLn(n).ent \mid n \in \polNodes \wedge \polLn(n).type=P\}$;
	\item $\Cg = \{\polLn(n).ent \mid n \in \polNodes \wedge \polLn(n).type=C\}$;
	\item $\Ag = \{\polLn(n).ent \mid n \in \polNodes \wedge \polLn(n).type=A\}$;
	\item $\Rg = \{\polLn(n).ent \mid n \in \polNodes \wedge \polLn(n).type=R\}$;
	\item $ \subseteq = \{(\polLn(n_{1}).ent, \polLn(n_{i}).ent) \mid \exists n_{2}, \dots, n_{i}, type(n_{1}, n_{2}, \dots, n_{i})=(\overrightarrow{CC}_{\actres})^{*} \} $;
	\item $ \oblOr = \{(\polLn(n_{1}).ent, \polLn(n_{i}).ent) \mid \exists n_{2}, \dots, n_{i}, type(n_{1}, n_{2}, \dots,n_{i})=(\overrightarrow{CC}_{\obl})^{*}\}$;
	\item $ \Eg = \{\polLn(n).ent \mid n \in \polNodes \wedge \polLn(n).type=\ev\} $;
	\item $ \GEg = \{\polLn(n).ent \mid n \in \polNodes \wedge \polLn(n).type=\evsch\} $;
	\item $ \Hg = \{ [\polLn(n_{1}).ent, \dots, \polLn(n_{i}).ent] \mid type(n_{1}, \dots, n_{i}) = (\overrightarrow{\ev\ev})^{*} \}$;
	\item $\PCAg = \{(\polLn(n_{1}).ent, \polLn(n_{2}).ent) \mid type(n_{1}, n_{2})=PC\}$;
	\item $\ARCAg = \{(a, r, \polLn(n_{1}).ent) \mid type(n_{1}, n_{2})=\textit{C\actres}^{A} \wedge \polLn(n_{2}).ent=(a, r)\}$;
	\item $\PARg = \{(\polLn(n_{1}).ent, a, r) \mid \exists n_{21}, \dots, n_{2i}, type(n_{1}, n_{21}, \dots, n_{2i}, n_{3})=PC, (\overrightarrow{CC}_{\actres})^{*},\textit{C\actres}^{A} \wedge \polLn(n_{3}).ent=(a, r)\}$;
	\item $\BARCAg = \{(a, r, \polLn(n_{1}).ent) \mid type(n_{1}, n_{2})=\textit{C\actres}^{B} \wedge \polLn(n_{2}).ent=(a, r)\}$;
	\item $\BARg = \{(\polLn(n_{1}).ent, a, r) \mid \exists n_{21}, \dots, n_{2i}, type(n_{1}, n_{21}, \dots, n_{2i}, n_{3})=PC, (\overleftarrow{CC}_{\actres})^{*},\textit{C\actres}^{B} \wedge \polLn(n_{3}).ent=(a, r)\}$;
	\item $\UNDETg = \{(\polLn(n_{1}).ent, \polLn(n_{2}).ent, \polLn(n_{3}).ent) \mid \polLn(n_{1}).type=P, \polLn(n_{2}).type=A, \polLn(n_{3}).type=R \}\setminus(\PARg \cup \BARg)$;
	\item $\OARCAg = \{(a, r, ge_{1}, ge_{2}, \polLn(n_{1}).ent) \mid type(n_{1}, n_{2})=\textit{C\obl} \wedge \polLn(n_{2}).ent=(a, r, ge_{1}, ge_{2})\}$;
	\item $\OPARg = \{(\polLn(n_{1}).ent, a, r, ge_{1}, ge_{2}) \mid \exists n_{21}, \dots, n_{2i}, type(n_{1}, n_{21}, \dots, n_{2i}, n_{3})=PC, (\overrightarrow{CC}_{\obl})^{*}, \textit{C\obl} \wedge \polLn(n_{3}).ent=(a, r, ge_{1}, ge_{2}) \} $;
	\item $\DPARg = \{ \polLn(n).ent \mid n \in \polNodes \wedge \polLn(n).type=\dut \} $;
	\item $\ETg = \{ (\polLn(n_{1}).ent, \polLn(n_{2i}).ent) \mid \exists n_{21}, \dots, n_{2i}, type(n_{1}, n_{21}, \dots, n_{2i})= \ev\evsch, (\overrightarrow{\evsch\evsch})^{*} \} $;
	\item $ \EIg = \{ (\polLn(n_{j}).ent, \polLn(n_{k}).ent, [\polLn(n_{1}).ent, \dots, \polLn(n_{i}).ent]) \mid type(n_{1}, \dots, n_{i}) = (\overrightarrow{\ev\ev})^{*} \wedge 1 \leq j < k \leq i \}$.
	
\end{itemize}

\end{definition}
\section{Representing Obligations in PORGY}
\label{sec:porgy}
We now describe how to implement the graphical representation of $\CBACO$ using PORGY. We start by describing our choices regarding the representation of the entities and the relations in the policies.

\subsubsection{Entities} Entities are represented as nodes of the graph, their $\textit{viewLabel}$ will be the same as $ent$ and the shape will be determined by the $type$. The number of ports is also determined by the $type$ of the entity. With the exception of edges representing the $ \subseteq $ and $\oblOr$ relations between categories,  the order of events in a history and the instantiation relation between event schemes, every other edge is undirected. Therefore, nodes of type $C$, $\ev$ and $\evsch$ will have three ports while the others only have one. The extra two ports (which have $\textit{viewLabel}$ \textbf{In} and \textbf{Out}) on these nodes are used to make the direction of the relation visually explicit.  These properties are summarized in Table~\ref{table:nodescbac}.

\begin{table}
  \begin{center}
    \begin{tabular}{|c|c|c|}
      \hline
      \textit{type} & \textit{viewShape} & \textit{number of ports} \\ \hline
      $P$ & Pentagon & 1 \\ \hline
      $C$ & Triangle & 3 \\ \hline
      $\actres$ & Hexagon & 1 \\ \hline
      $A$ & Square & 1 \\ \hline
      $R$ & Diamond & 1 \\ \hline
      $\obl$ & Hexagon & 1 \\ \hline
      $\dut$ & Hexagon & 1 \\ \hline
      $\ev$ & Circle & 3 \\ \hline
      $\evsch$ & Ring & 3 \\ \hline
    \end{tabular}
  \end{center}
  \caption{Properties of nodes in a $\CBACO$ representation}
  \label{table:nodescbac}
  \vspace{-0.2in}
\end{table}
\vspace{-0.2in}
\subsubsection{Relations} Relations are represented as edges of the graph. The \textit{viewColor} of edges is gray except for edges of type $\textit{C\actres}$, $\obl\evsch$ and $\dut\ev$.  Edges of type $\textit{C\actres}$ are green if its \textit{auth} is $A$ and red if it is $B$ and edges of type $\obl\evsch$ (resp. $\dut\ev$) will be green if $ge$ (resp. $ev$) is $i$ and red if it is $f$. Edges of type $CC$, $\evsch\evsch$ and $\ev\ev$ will connect to the extra ports of nodes with $type$ $C$, $\evsch$ and $\ev$, respectively, connecting to the \textbf{In} port if the $ent$ of the node is in the $\textit{target}$ of the edge or to the \textbf{Out} port otherwise. 

\subsubsection{Color coding} In order to distinguish categories that belong to the permission/prohibition subgraph from the ones that belong to the obligation subgraph and the ones that are used in both, a color coding is used. This color coding was then extended to every type of node in order to make the graph more informative. The coding is simple, if a node belongs solely to the permission/prohibition subgraph then its color is yellow, if it belongs solely to the obligation subgraph then its color is blue and if it belongs to both then its color is green. We use the color also to make explicit which of the event nodes has its $now$ set to $\top$ by setting its color to a lighter blue. 
\begin{example} \label{ex:cbaco_rep}
In Figure~\ref{fig:ex_cbaco_rep} we have the graphical representation of a $\CBACO$ policy such that:

\begin{itemize}
	\item $ \P = \{ \textit{J. Dorian}, \textit{C. Tuck} \} $;
	\item $ \C = \{ \textit{Dr(J. Lewis)}, \textit{Dr(F. Mason)} \} $;
	\item $ \A = \{ \textit{Read}, \textit{Declare} \} $;
	\item $ \R = \{ \textit{Rec(J. Lewis)}, \textit{Rec(F. Mason)}, \textit{Admin-log} \} $;
	\item $ \subseteq = \emptyset $;
	\item $ \oblOr = \emptyset $;
	\item $ \E = \{ \{ \act = \textit{Read}, \subj = \textit{C. Tuck}, \obj = \textit{Rec(J. Lewis)}, \evtime=120 \}, \{ \act = \textit{Declare}, \subj = \textit{C. Tuck}, \obj = \textit{Admin-log}, \evtime=200 \} \} $;
	\item $ \GE = \{ \textit{gen\_read}[\textit{J. Dorian}, \textit{F. Mason}], \textit{gen\_read}[\textit{C. Tuck}, \textit{J. Lewis}] \} $;
	\item $ \H = \{ [  \{ \act = \textit{Read}, \subj = \textit{C. Tuck}, \obj = \textit{Rec(J. Lewis)}, \evtime=120 \},  \{ \act = \textit{Declare}, \subj = \textit{C. Tuck}, \obj = \textit{Admin-log}, \evtime=200 \} ] \} $;
	\item $ \PCA = \{ (\textit{J. Dorian}, \textit{Dr(J. Lewis)}), (\textit{C. Tuck}, \textit{Dr(F. Mason)}) \} $;
	\item $ \ARCA = \{ (\textit{Read}, \textit{Rec(F. Mason)}, \textit{Dr(F. Mason)}), (\textit{Read}, \textit{Rec(J. Lewis)}, \textit{Dr(J. Lewis)}) \} $;
	\item $ \PAR = \{ (\textit{J. Dorian}, \textit{Read}, \textit{Rec(J. Lewis)}), (\textit{C. Tuck}, \textit{Read}, \textit{Rec(F. Mason)}) \} $;
	\item $ \BARCA = \emptyset $;
	\item $ \BAR = \emptyset $;
	\item $ \UNDET = \{ ( \textit{C. Tuck}, \textit{Read},\textit{Rec(J. Lewis)}), ( \textit{J. Dorian}, \textit{Declare},\textit{Rec(J. Lewis)}),\\ ( \textit{C. Tuck}, \textit{Declare},\textit{Rec(J. Lewis)}), ( \textit{J. Dorian}, \textit{Read},\textit{Rec(F. Mason)}), \\( \textit{J. Dorian}, \textit{Declare},\textit{Rec(F. Mason)}), ( \textit{C. Tuck}, \textit{Declare},\textit{Rec(F. Mason)}), \\( \textit{J. Dorian}, \textit{Read},\textit{Admin-log}), ( \textit{C. Tuck}, \textit{Read},\textit{Admin-log}),\\ ( \textit{J. Dorian}, \textit{Declare},\textit{Admin-log}), ( \textit{C. Tuck}, \textit{Declare},\textit{Admin-log}) \} $;
	\item $ \OARCA = \{ ( \textit{Declare}, \textit{Admin-log}, \textit{gen\_read}[\textit{J. Dorian}, \textit{F. Mason}], \bot, \textit{Dr(J. Lewis)} ),\\ ( \textit{Declare}, \textit{Admin-log}, \textit{gen\_read}[\textit{C. Tuck}, \textit{J. Lewis}], \bot, \textit{Dr(F. Mason)} ) \} $;
	\item $ \OPAR = \{ ( \textit{J. Dorian}, \textit{Declare}, \textit{Admin-log}, \textit{gen\_read}[\textit{J. Dorian}, \textit{F. Mason}], \bot ),\\ ( \textit{C. Tuck}, \textit{Declare}, \textit{Admin-log}, \textit{gen\_read}[\textit{C. Tuck}, \textit{J. Lewis}], \bot ) \} $;
	\item $ \DPAR = \{ ( \textit{C. Tuck}, \textit{Declare}, \textit{Admin-log}, \{ \act = \textit{Read}, \subj = \textit{C. Tuck}, \obj = \textit{Rec(J. Lewis)}, \evtime=120 \}, \bot ) \} $;
	\item $ \ET = \{ ( \{ \act = \textit{Read}, \subj = \textit{C. Tuck}, \obj = \textit{Rec(J. Lewis)} \},\\ \textit{gen\_read}[\textit{C. Tuck}, \textit{J. Lewis}] ) \} $;
	\item $ \EI = \{ ( \{ \act = \textit{Read}, \subj = \textit{C. Tuck}, \obj = \textit{Rec(J. Lewis)}, \evtime = 120 \}, \{ \act = \textit{Declare}, \subj = \textit{C. Tuck}, \obj = \textit{Admin-log}, \evtime = 200 \}, [ \{ \act = \textit{Read}, \subj = \textit{C. Tuck}, \obj = \textit{Rec(J. Lewis)}, \evtime = 120 \},\\ \{ \act = \textit{Declare}, \subj = \textit{C. Tuck}, \obj = \textit{Admin-log}, \evtime = 200 \} ] ) \} $.
\end{itemize}
Note that the duty is not represented in the graph, this is due to the fact that we use rewrite rules and the strategy language to instantiate and update the state of duties in a history of events. This process is shown in Example~\ref{ex:duty}.

\begin{figure}
\centering
\includegraphics[width=\linewidth]{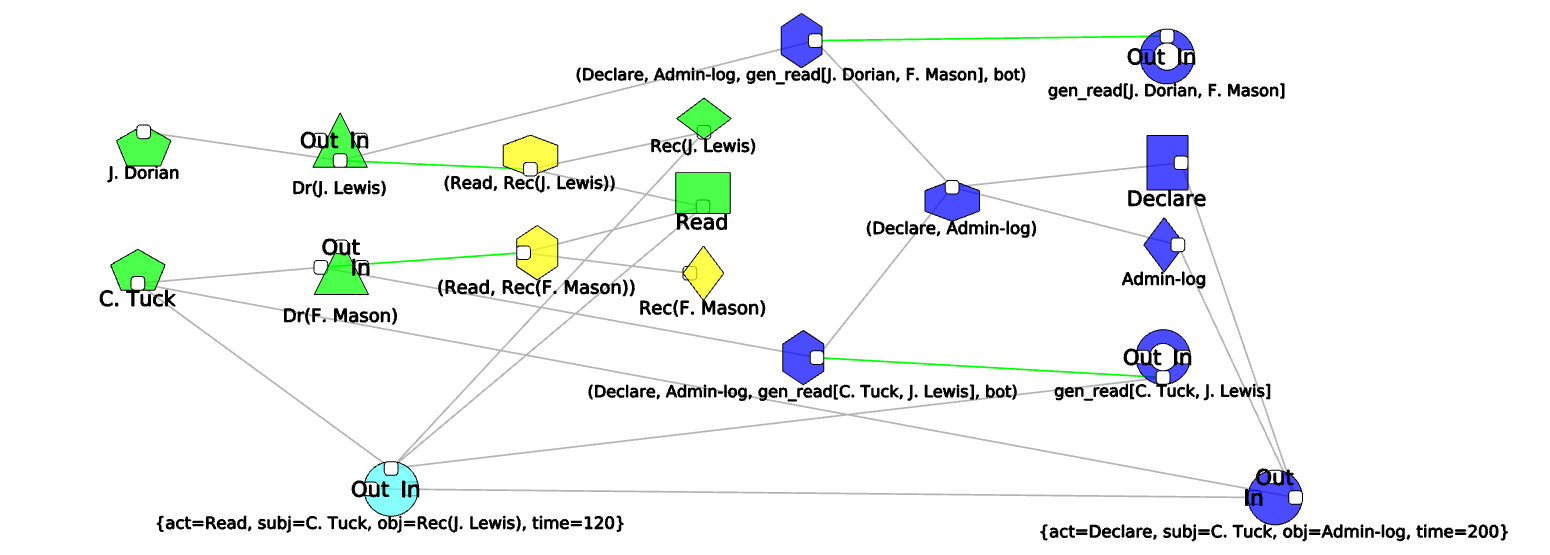}
\caption{$\CBACO$ policy representation}
\label{fig:ex_cbaco_rep}
\end{figure}
\end{example}

\section{The Dynamics of Obligations in PORGY}
\label{sec:dyno}
A key point of $\CBACO$\ policies is that the assignment of principals, permissions, prohibitions and obligations to categories is not static, it depends on the system state. We will now see how we can use rewrite rules and the strategy language to ensure that relations on the graph are correct. These rules fall into two main groups, the rules that are general to every \CBACO\ policy  and  will  not  change  the  policy  itself  (the  ones  associated  with  queries  and obligations), and the rules that are scenario specific and will change the policy. 
\vspace{-0.1in}
\subsection{Scenario-specific rewrite rules}
This  category  covers  a  wide  variety  of  rules that are dependant on the policy.  Their purpose is to represent the dynamics specific to the policy: ways of creating/removing entities, updating relations or constraints of the policy.  They can be triggered either by events or by changes in the state.
\begin{figure}[ht]
    \centering
{\small
\begin{verbatim}
		setBan(all(property(crtGraph,node,type=="P")));
		while(not(isEmpty(crtBan)))do(
		  setPos(one(crtBan));
		  setBan(all(crtBan\crtPos));
		  setPos(all(crtPos[cup]ngb(crtPos,edge,type=="PC")));
		  while(one(auxPC))do(
		    repeat(one(auxPC));
		    setPos(all(
		    property(crtPos,node,type=="P")[cup]property(crtBan,node,type=="C")
		    ));
		    setBan(all(property(crtBan,node,type=="P")))
		))
\end{verbatim}
}
    \caption{A strategy for applying rule \textsf{auxPC}}
    \label{fig:script}
\end{figure}
\vspace{-0.2in}
\subsection{General rewrite rules}
 Inside the scope of these rewrite rules we have: auxiliary rules that extract information used in other rules; rules related to the treatment of obligations; and rules related to visualisation and queries.
\subsubsection{Auxiliary rewrite rules} Their  purpose  is  to make implicit relations explicit, nonetheless, they will be invisible so they can be used in rewrite rules but will not appear in the graph visualization.  Note that a policy graph does not contain redundant edges. However, redundant edges are often necessary to establish relations (for example, principal-category relations in the presence of the $\subseteq$ relation on categories). All the auxiliary edges  will have attribute \textit{aux} set to \textit{true}.

The rule in Figure~\ref{fig:ex1}, named \textsf{auxPC}, is an example of a port-graph rule used to build auxiliary edges. Using the strategy language of PORGY this rule can be used exhaustively, until no more new edges of type $PC$ are created (see Figure~\ref{fig:script} for a script implementing this strategy).

\subsubsection{Obligation rewrite rules}
Obligation rewrite rules will concern the dynamics of duties and its states. In this category of rules we have rules to deal with instantiating duties and rules for every possible update of the state of a duty. Also in this category we put the rules concerning the iteration of events, this is due to the fact that, although they could also be necessary for the scenario-specific rules, they are central to the calculation of duties and their states.
\begin{example}\label{ex:duty}
\begin{figure}[ht]
\centering
\begin{subfigure}{\linewidth}
\includegraphics[width=0.9\linewidth]{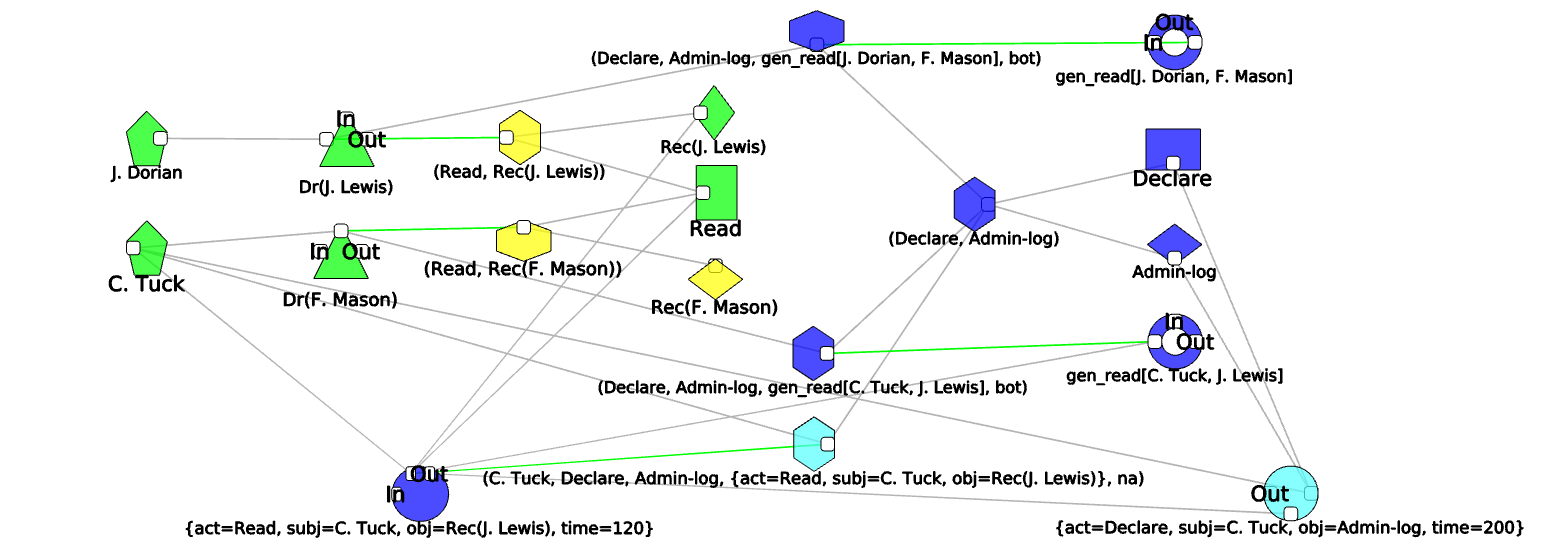} 
\caption{After first event is processed}
\label{fig:ex_cbaco_rep1}
\end{subfigure}
\begin{subfigure}{\linewidth}
\includegraphics[width=0.9\linewidth]{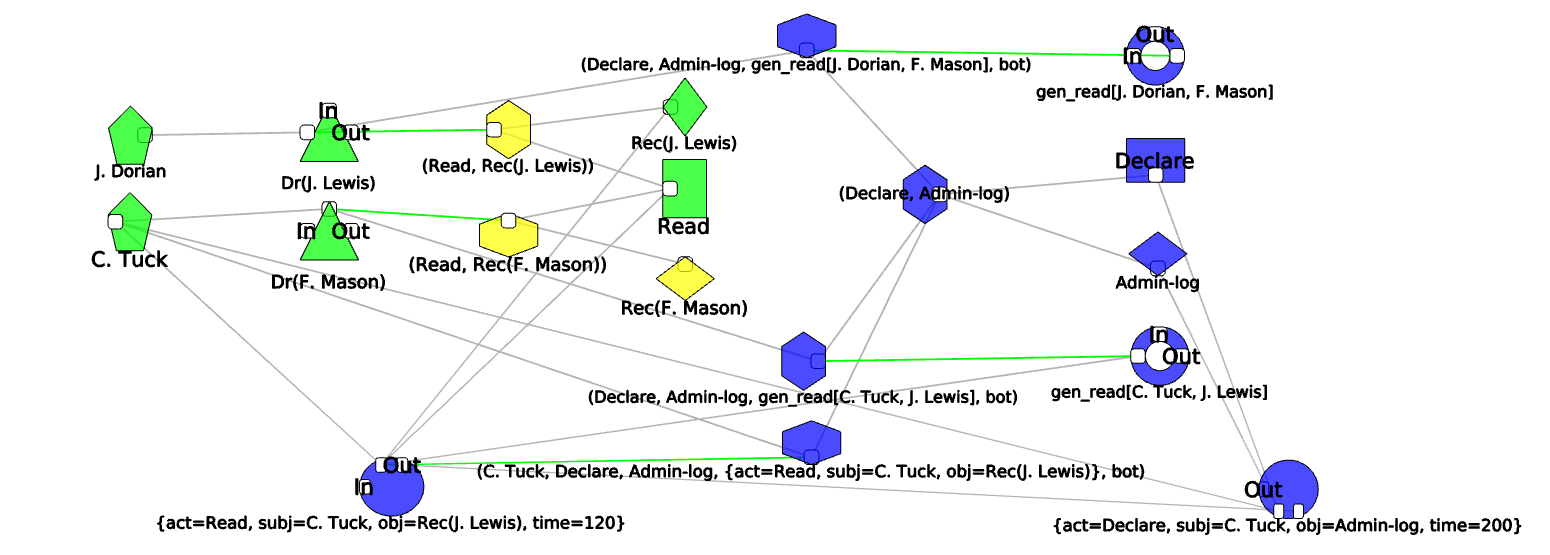}
\caption{After both events are processed}
\label{fig:ex_cbaco_rep2}
\end{subfigure}
\caption{$\CBACO$ policy representations}
\end{figure}
Recall the graphical representation of a $\CBACO$ policy in Figure~\ref{fig:ex_cbaco_rep}. 
As we can see the active event node is connected to an event scheme node, which is connected to an obligation node by a green edge meaning it will trigger a duty. This is shown in Figure~\ref{fig:ex_cbaco_rep1}. The only change in the graph is the addition of this duty node and respective edges.

Finally, the active event in Figure~\ref{fig:ex_cbaco_rep1} represents the action on the resource by the subject needed to fulfill the duty, and so the final graph is shown in Figure~\ref{fig:ex_cbaco_rep2} where we get the representation of the duty present in the policy.
\end{example}

\subsubsection{Visualisation and queries}
Queries rewrite rules concern possible queries that can be made by a system manager. They will rely on the rules mentioned above (auxiliary and obligation) to get the requested answer coupled with visualization rules to show said answer. Visualization rules are just rules to hide or show specific nodes or edges, based on their attributes, that can be used to get different views of the policy graph by using certain subsets of them. 

\section{Related Work}
\label{sec:rw}
The use of graph based languages to model and analyse policies has mostly been developed in the context of RBAC~\cite{Baldwin90,NyanchamaO99,KochMP00,KochMP02,ThionC06}. Most work done on $\CBAC$ policies uses only textual languages and focuses on the expressiveness of the model, the analysis of policies and techniques that can be used to enforce policies\cite{Barker09,BertolissiF11,BertolissiF14,AliF14}. Graph based models for the analysis of CBAC policies, were introduced in~\cite{AlvesF15,AlvesF17}, dealing with both permissions and prohibitions, but neither addresses $\CBACO$ policies.

In \cite{KochMP00,KochMP02}, inspired by Nyanchama and Osborn in \cite{NyanchamaO99}, Koch et al. use directed graphs to analyse $\RBAC$ policies using a graph model similar to Baldwin's privilege graphs \cite{Baldwin90}. One key aspect of \cite{KochMP00,KochMP02} is the use of graph transformations to model administrative operations on $\RBAC$ policies. This focus on administrative operations comes from the fact that $\RBAC$ policies are static, whereas in $\CBAC$ the same operations are dependant on the system state and are carried out without the need for an administrator to intervene. Using our model we could represent the $\RBAC$ policies in \cite{KochMP00,KochMP02} since roles are a particular case of category; however, the graphs in \cite{KochMP00,KochMP02} also deal with sessions. In this paper we have not dealt with sessions, but since the notion of session in $\RBAC$ is similar to the notion of session in $\CBAC$ it would be easy to adapt the formalism used in \cite{KochMP00,KochMP02} to our model. In \cite{KochMP01a,KochMP01} a framework for the description of access control models using graph transformations is described. With each type of access control model is associated a typegraph for its graph based representation. Graph transformations are used to describe both the ways that the policy can evolve as well as positive and negative constraints. One of the focus of the work is the integration of policies that use different access control models. As the $\CBAC$ metamodel has been shown to subsume many of the most well known access control models it makes it a good candidate for this purpose.

In \cite{KochP03} three modeling notations (viz. UML, Alloy and Graph Transformations) are compared for the task of specification and verification of RBAC models. The focus of the paper is on the evolution and constraints of the policies and, with it, the possible violation of constraints that may arise. Although we have not dealt with constraints in our implementation it would be easy to integrate them.
In \cite{ThionC06} conceptual graphs and conceptual graph rules are used to represent and reason on RBAC policies. Conceptual graphs are just labelled hypergraphs, meaning edges (called hyperedges) can represent any n-ary relationship between nodes. Rules are used to derive authorisation information from the policy graph and to formulate constraints. One interesting aspect is the use of backwards chaining to find inconsistencies in the policy; starting from a rule defining a constraint they work their way backwards (using the rules of information derivation in the opposite direction) to find if the policy could violate said constraint. While such mechanism is not possible in PORGY we could still check for the same constraint violations.

Miró \cite{HeydonMTWZ90} is a set of languages and tools for visual specification of file system security inspired by Harel's work on higraphs \cite{Harel88}. Higraphs are a visual formalism of topological nature that combine Euler/Venn diagrams with hypergraphs. Instead of nodes we have blobs that can contain/intersect other blobs and can be connected by hyperedges making it a powerful formalism to represent a wide variety of complex systems. In the case of Miró, blobs are used to represent sets of users and files, and binary labeled directed edges describe the granting/denial of access rights or, in some cases (constraint language), if a blob is or is not inside another blob. There are two languages, an instance language that describes the access rights of users to files, and the constraint language that restricts the set of allowed instance pictures. The blobs representing sets of users are a kind of categorisation and as so this part of the representation is easily translated into our model. The blobs representing sets of files is something we cannot directly translate to our model as it is right now, but could easily be integrated as a partial ordering of resources such as with categories. As stated edges can be used to represent denial of access rights as in our model but, in contrast to our model, these denials propagate from the more general blobs to the less general as with granting. Furthermore, the lack of an edge granting some user access to a file is interpreted as denial of access and there can be multiple edges that apply to the same triple user, action, file, in which case a predominant edge is searched, if there is no such edge it classifies as an ambiguity (in our model such ambiguities are not allowed by definition). Because of these differences, mapping a higraph in the Miró languages to a $\CBACO$ policy graph is not trivial, nonetheless, policies described using the Miró languages could be represented using our model.

LaSCO \cite{Hoagland00} is a policy language based on graphs. The graphs are labelled; nodes represent objects (users, classes, resources, etc) and directed edges represent actions (events in the LaSCO terminology). A policy graph in LaSCO describes both the system state to which it applies and the constraints on that state for it to be valid; this is done with the use of labels (predicates in the LaSCO terminology). These labels can be seen as type of categorisation and so we can map these policy graphs into policies in our model.

In \cite{TidswellP01} graphs are used to define a framework for the representation of various access control models. Based on this framework a visual language for the definition of constraints is also defined. Another focus of the paper is on triggers, events that change the state of the policy or system. With some adaptation in some cases, every model, constraint and trigger discussed could be translated into our model.
\section{Evaluation and Future Work}
\label{sec:concl}
This paper presents a graph-based metamodel that deals with obligations in access control, together with its implementation in the PORGY framework. This is a prototipal implementation, for which we have not yet develop performance evaluations. Nonetheless, we will give a brief evaluation of our framework by highlighting its main advantages:
\begin{itemize}
\item The proposed model can deal with obligations in a comprehensive manner, as was shown by the comparison with the related work in the previous section.
\item The strategy driven language of PORGY is powerful enough to deal with the dynamics aspects of the proposed model. Using the rewriting language it is possible to simulate changes in a system, which are presented in a derivation tree where each node can be inspected (see Figure~\ref{fig:trace}).
\begin{figure}
\centering
\includegraphics[width=0.9\linewidth]{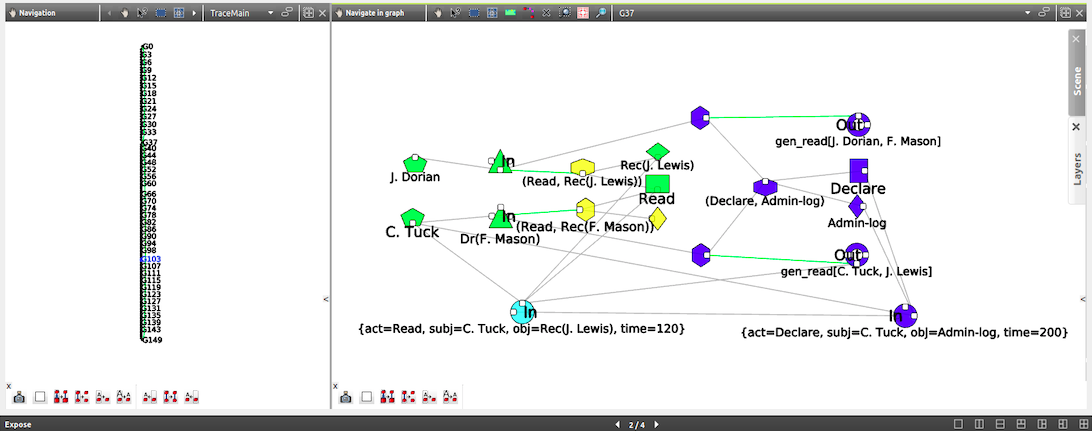}
\caption{Tracing a strategy}
\label{fig:trace}
\end{figure}
\item Rewriting systems have the advantages of being multi-paradigm and having a well-developed theory. For example, the rewrite-based operational semantics of the \CBACO\ metamodel~\cite{AlvesDF15}, allows for the  analysis of compatibility between authorisations and obligations, and we have used PORGY's rewriting language to evaluate the state of duties.
\item PORGY is being actively developed, which opens the door for further improvement and the PORGY team is open to develop domain specific versions of the framework, with tailored visualisation applications.
\item The PORGY tool is built on top of Tulip~\cite{Auber2017}, an information visualization framework dedicated to the analysis and visualization of relational data, and both are released under the LGPL licence. The Tulip framework can be used to import/export information from PORGY, which is a positive aspect for integration with other prototyping and analysis tools.
\end{itemize}
\begin{figure}[ht]
    \centering
    \includegraphics[width=0.9\linewidth]{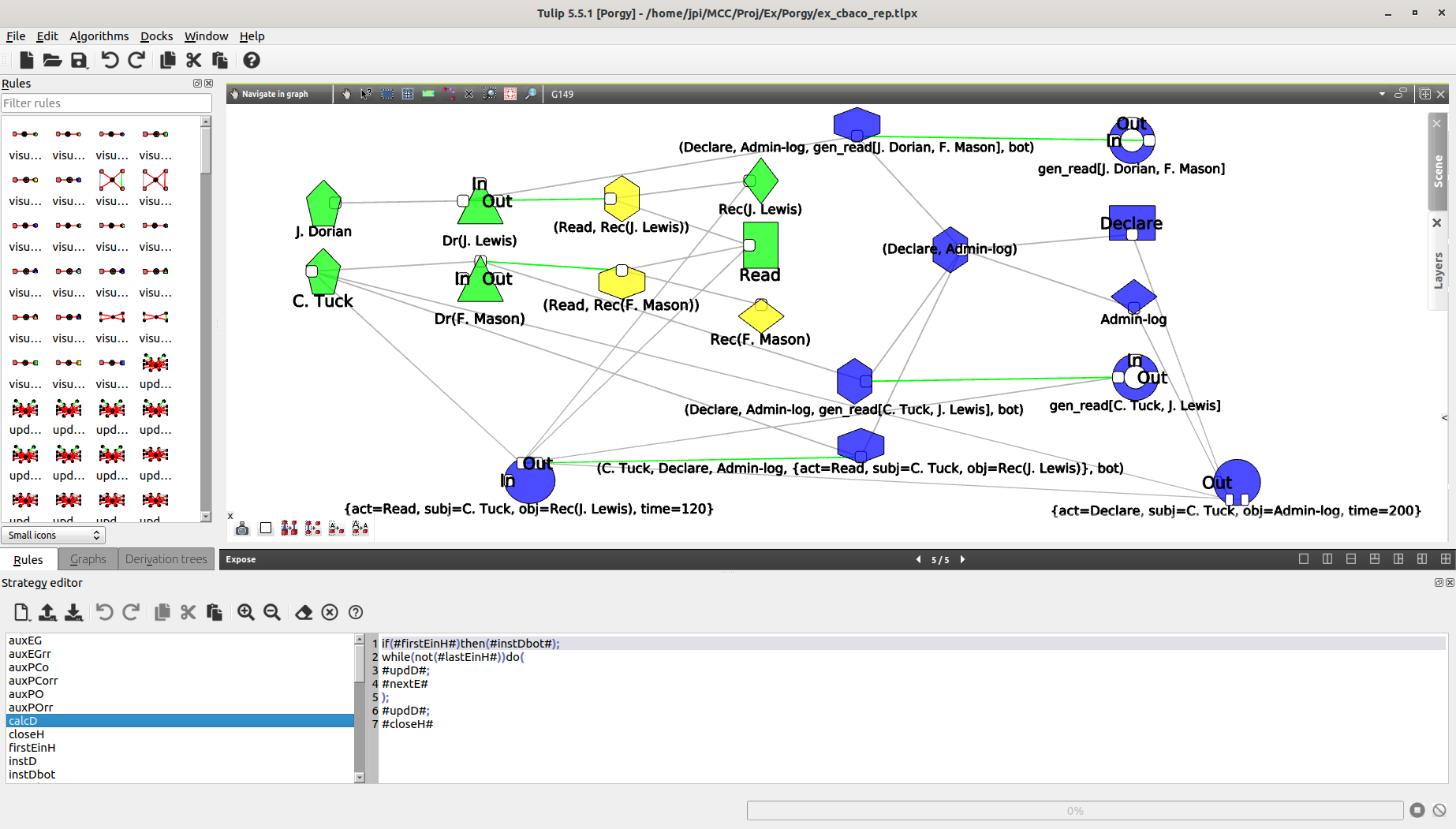}
    \caption{PORGY's interface}
    \label{fig:interface}
    \end{figure}
We now point out some less positive aspects of our framework that we would like to further develop:
\begin{itemize}
    \item Presently there is a huge repetition of rules that result from the need to define the rules for all the nodes of the same type that may differ in their attributes (for example, its colour). Although this leads to a rather cumbersome translation of the policy graphs into PORGY, this aspect could be significantly improved by automating the generation of duplicated graphs/rules. This could also be eased by the use of higher-order notions of port-graphs that could facilitate the abstraction of patterns in the policy graphs, leading to more efficient ways of writing rules that depend on information on nodes that are not directly connected.
    \item PORGY framework has a well-developed interface, where graphs and rules to be constructed, strategies can be edited, applied and its trace can be followed (see Figure~\ref{fig:interface}). However, policy administrators could benefit from the existence of user-friendly Domain Specific Languages (DSLs), that could be translated into our modelling languages or other programming languages for integration with existing policy analysis tools. 
    \item  We would like to use the prototyping and analysis capabilities of rewriting to analyse additional properties related to obligations, such as accountability.
\end{itemize}

\bibliographystyle{abbrv}
\bibliography{main}
\end{document}